\documentstyle[amssymb,floats,aps,prb,epsf]{revtex}
\epsfclipon 
\begin{document}
\twocolumn[\hsize\textwidth\columnwidth\hsize\csname
@twocolumnfalse\endcsname 

\draft

\title{Spin Dynamics for the $t$-$J$ Model}
\author{Zhongbing Huang$^{2}$ and Shiping Feng$^{1,2,3}$}
\address{$^{1}$CCAST (World Laboratory) P. O. Box 8730, Beijing
100080, China \\
$^{2*}$Department of Physics, Beijing Normal University, Beijing
 100875, China \\
$^{3}$National Laboratory of Superconductivity, Academia Sinica,
 Beijing 100080, China }

\maketitle
\begin{abstract}
The spin dynamics at the finite temperature for the $t$-$J$ model
in the underdoped and optimal doped regimes is studied within the
fermion-spin theory. It is shown that the dynamical spin structure
factor spectrum at the antiferromagnetic wave vector $Q=(\pi,\pi)$
are separated as low- and high-frequency parts, respectively, but
the high-frequency part is suppressed in the dynamical
susceptibility spectrum $\chi^{''}(Q,\omega)$, while the
low-frequency part is the temperature dependent, which are in
qualitative agreement with the experiments and numerical
simulations.
\end{abstract}
\pacs{71.27.+a, 74.72.-h, 76.60.-k}
]
\bigskip

\narrowtext

The $t$-$J$ model provides one of the most interesting and simplest
models to study the physics of the doped Mott insulator \cite{n1}.
It was originally introduced as an effective Hamiltonian of the
Hubbard model in the strong coupling regime \cite{n2}, where the
on-site Coulomb repulsion $U$ is very large as compared with the
electron hopping energy $t$, and in this case the electrons become
strongly correlated to avoid the double occupancy, i.e.,
$\sum_{\sigma}C_{i\sigma}^{+}C_{i\sigma}^{-}\leq 1$. The interest
in the two-dimensional (2D) $t$-$J$ model was stimulated by many
researchers' suggestion that it contains the relevant physics of
the copper oxide superconductors \cite{n1}.

During the past ten years, consistent experiments picture of the
magnetic properties of the copper oxide materials have been emerged
\cite{n3,n4}: the undoped copper oxide materials are
antiferromagnetic Mott insulators, and upon doping with holes in the
copper oxide sheets, the antiferromagnetic long-range-order (AFLRO)
is replaced by short-range spin correlations, and the dynamic
antiferromagnetic correlations persist even into the superconducting
state. A series of the neutron-scattering study \cite{n5,n6} of the
copper oxide materials shows that there is an anomalous temperature
dependence of the spin fluctuations near the antiferromagnetic zone
center in the underdoped and optimal doped regimes, which has a much
larger effect on the spin dynamics and leads to the unusual
temperature dependence of the spin-lattice relaxation time. On the
theoretical side, the spin dynamics in the underdoped and optimal
doped regimes has been studied within the phenomenological model of
antiferromagnetic correlated spins \cite{n7}, and the framework of
the phenomenological marginal Fermi-liquid theory \cite{n8}.
Moreover, the most reliable results of the spin dynamics have been
obtained by the numerical simulations \cite{n9} and high-temperature
series expansion \cite{n10} based on some strongly correlated
models. It is believed that the role played by the magnetism,
particularly the nature of spin fluctuations, is the central issue
of the copper oxide materials. In this paper, we employ the $t$-$J$
model to study the spin dynamics of the copper oxide materials in
the underdoped and optimal doped regimes within the fermion-spin
theory \cite{n11,n12}, and show that our theoretical results are
qualitative consistent with the experiments \cite{n5} and numerical
simulations \cite{n9}.

In order to account for the real experiments under the $t$-$J$
model, the crucial requirement is to impose the electron on-site
local constraint for a proper understanding of the physics of the
copper oxide materials \cite{n121}. For incorporating the local
constraint, the fermion-spin theory based on the charge-spin
separation has been proposed \cite{n11}. In this approach, the
constrained electron operators are decoupled as
$C_{i\uparrow}=h_{i}^{+}S_{i}^{-}$ and
$C_{i\downarrow}=h_{i}^{+}S_{i}^{+}$, with the spinless fermion
operator $h_{i}$ keeps track of the charge (holon), while the
pseudospin operator $S_{i}$ keeps track of the spin (spinon). The
advantage of the fermion-spin theory is that the electron on-site
local constraint for the single occupancy is satisfied even in the
mean-field approximation (MFA). Within the fermion-spin theory, the
mean-field theory \cite{n12} in the underdoped and optimal doped
regimes has been developed, where the mean-field order parameters
are defined as $\chi=\langle S_{i}^{+}S_{i+\eta }^{-}\rangle =
\langle S_{i}^{-}S_{i+\eta}^{+}\rangle$, $\chi_{z}=\langle
S_{i}^{z}S_{i+\eta }^{z}\rangle$,
$C=(1/Z^{2})\sum_{\eta ,\eta ^{\prime }}\langle S_{i+\eta }^{+}
S_{i+\eta ^{\prime}}^{-} \rangle$, $C_{z}=(1/Z^{2})
\sum_{\eta ,\eta ^{\prime }}\langle S_{i+\eta }^{z}
S_{i+\eta ^{\prime }}^{z}\rangle$, and $\phi =\langle
h_{i}^{\dagger}h_{i+\eta }\rangle$, with
$\hat{\eta }=\pm \hat{x},\pm \hat{y}$, and $Z$ is the number of
nearest neighbor sites. This mean-field theory \cite{n12} has been
applied to study the electron spectrum, electron dispersion, and
electron density of state of the copper oxide materials, and the
results are in agreement with the experiments and numerical
simulations. In the fermion-spin representation, the $t$-$J$ model
is written as $H=H_{t}+H_{J}$ with
\begin{mathletters}
\begin{eqnarray}
H_t&=&-t\sum_{i\eta}h_{i}h_{i+\eta}^{\dagger}(S_{i}^{+}
S_{i+\eta}^{-}+S_{i}^{-}S_{i+\eta}^{+})+h.c. \nonumber \\
&+&\mu \sum_{i}
h_{i}^{\dagger}h_{i} , \\
H_J&=&J_{eff}\sum_{i\eta}[{1 \over 2}(S_{i}^{+}
S_{i+\eta}^{-}+S_{i}^{-}S_{i+\eta}^{+})+S_{i}^{z}
S_{i+\eta}^{z}],
\end{eqnarray}
\end{mathletters}
where $J_{eff}=J[(1-\delta)^{2}-\phi ^{2}]$, and $\mu$ is the
chemical potential. Based on the Ioffe-Larkin combination rule
\cite{n13}, the charge dynamics of the copper oxide materials
in the underdoped and optimal doped regimes has been discussed
\cite{n14} by considering the charge fluctuations around the
mean-field solution. However, the spin fluctuations couple only
to spinons and therefore no composition law is required
\cite{n13} in discussing the spin dynamics, but the effect of
holons still is considered through the holon's order parameter
$\phi$ entering in the spinon propagator. For discussing the spin
dynamics, we need to calculate the second-order correction for
the spinon by going beyond MFA. The second-order spinon
self-energy diagram from the holon pair bubble is shown in Fig. 1.
The mean-field spinon Green's functions
$D^{(0)}(k,\omega)$ and $D_{z}^{(0)}(k,\omega)$, and mean-field
holon Green's function $g^{(0)}(k,\omega)$ have been given
in Ref. \cite{n12}. Since the spinon operator obey the Pauli
algebra, it is needed to map the spinon operator into the
CP$^{1}$ fermion representation or the spinless-fermion
representation in terms of the $2D$ Jordan-Wigner transformation
\cite{n15} for the formal many particle perturbation expansion.
After such formal expansion, the spinon Green's function in
the spinon self-energy diagram shown in Fig. 1 is replaced by the
mean-field spinon Green's function $D^{(0)}(k,\omega)$, then the
second-order spinon self-energy is evaluated as,
\begin{eqnarray}
\Sigma_{s}^{(2)}(k,\omega)&=&-(Zt)^{2}{1\over N^2}\sum_{pp'}
(\gamma_{k-p}+\gamma_{p'+p+k})^{2}B_{k+p'} \nonumber \\
&\times& \left ({F_{1}(k,p,p')\over \omega +\xi_{p+p'}-\xi_{p}+
\omega_{k+p'}+i0^{+}} \right. \nonumber \\
&-&\left. {F_{2}(k,p,p')\over \omega +\xi_{p+p'}-\xi_{p}-
\omega_{k+p'}+i0^{+}} \right ) ,
\end{eqnarray}
where $F_{1}(k,p,p')=n_{F}(\xi_{p+p'})[1-n_{F}(\xi_{p})]+[1+n_{B}
(\omega_{k+p'})] [n_{F}(\xi_{p})-n_{F}(\xi_{p+p'})]$,
$F_{2}(k,p,p')=n_{F}(\xi_{p+p'})[1-n_{F}(\xi_{p})]-n_{B}
(\omega_{k+p'})[n_{F}(\xi_{p})-n_{F}(\xi_{p+p'})]$
$n_{F}(\xi_{k})$ and $n_{B}(\omega_{k})$ are the Fermi and
Bose distribution functions, respectively,
$\gamma_{{\bf k}}=(1/Z)\sum_{\eta}e^{i{\bf k}\cdot\hat{\eta}}$,
$B_{k}={ZJ_{eff}[(2\epsilon\chi_{z}+\chi)\gamma_{k}-
(\epsilon\chi+2\chi_{z})]/\omega (k)}$, $\epsilon=1+2t\phi/J_{eff}$,
and the mean-field holon excitation spectrum $\xi_{k}$ and
mean-field spinon excitation spectrum $\omega_{k}$ are given in
Ref. \cite{n12}. In this case, the full spinon Green's function is
obtained as
$D^{-1}(k,\omega)=D^{(0)-1}(k,\omega)-\sum_{s}^{(2)}(k,\omega)$.

\begin{figure}[prb]
\epsfxsize=3.0in\centerline{\epsffile{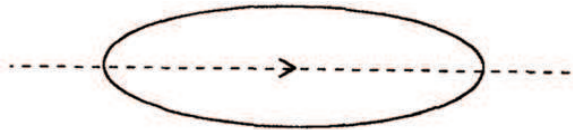}}
\caption{The spinon's second-order self-energy diagram. The solid
and dashed lines correspond to the holon and spinon propagators,
respectively.}
\end{figure}

\begin{figure}[prb]
\epsfxsize=3.0in\centerline{\epsffile{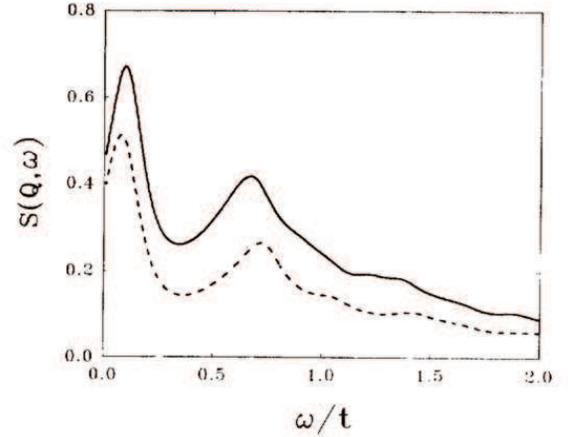}}
\caption{The dynamical spin structure factor spectra $S(Q,\omega)$
in the temperature $T=0.2J$ for the doping $\delta=0.06$ (solid
line) and $\delta=0.15$ (dashed line).}
\end{figure}

\begin{figure}[prb]
\epsfxsize=3.0in\centerline{\epsffile{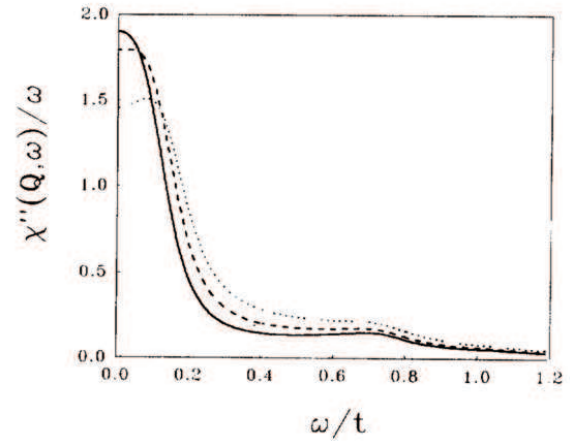}}
\caption{The dynamical spin susceptibility spectra
$\chi^{''}(Q,\omega)/\omega$ at the doping $\delta=0.15$ for the
temperature $T=0.2J$ (solid line), $T=0.3J$ (dashed line), and
$T=0.45J$ (dotted line).}
\end{figure}

As manifestation of the spin dynamics, the dynamical spin
structure factor $S(k,\omega)$ and dynamical susceptibility
$\chi (k,\omega)$ are given as
\begin{eqnarray}
S(k,\omega)&=&Re\int^{\infty}_{0}dt e^{i\omega t}\langle
S^{+}_{k}(t)S^{-}_{k}(0)\rangle \nonumber \\
&=&{2{\rm Im}D(k,\omega)\over(1-e^{-\beta \omega})},
\end{eqnarray}
and
\begin{eqnarray}
\chi^{\prime\prime}(k,\omega)&=&(1-e^{-\beta \omega})S(k,\omega)
\nonumber \\
&=&2{\rm Im}D(k,\omega),
\end{eqnarray}
respectively. Although the detailed magnetic properties depend
on the sample preparation method in the experiments as well as the
precise value and homogeneity of the oxygen content, the anomalous
temperature dependence of the spin fluctuations near the
antiferromagnetic wave vector $Q=(\pi,\pi)$ seems common
\cite{n5,n6,n7}. We have performed the numerical calculation for
the dynamical structure factor (3) and dynamical susceptibility
(4). The result of the $S(Q,\omega)$ spectra at the doping
$\delta =0.06$ (solid line) and $\delta =0.15$ (dashed line) with
the temperature $T=0.2J$ are plotted in Fig. 2. It is shown that
in the underdoped and optimal doped regimes there are the
coexistence of the low- and high-frequency fluctuations in the
$S(Q,\omega)$ spectra, the excitations are remarkably sharp, and
the spectra are changed with dopings, which is consistent with
the experiments \cite{n5} and numerical simulation \cite{n9}.
The low-frequency peak in $Q$ point is due to the antiferromagnetic
fluctuations, which will be in existence even in the undoped case,
and dominate dynamical susceptibility and the neutron-scattering
processes, while the high-frequency peak may come from the
contribution of the free-fermion-like component of the systems,
which induces the main effect to the large extent the static spin
correlation. In correspondence with the $S(Q,\omega)$ spectra,
the numerical results of the dynamical susceptibility spectra
$\chi^{''}(Q,\omega)$ at the doping $\delta =0.15$ for the
temperature $T=0.2J$ (solid line), $T=0.3J$ (dashed line),
$T=0.45J$ (dotted line) are plotted in Fig. 3. Comparing with the
Fig. 2, it is shown that although the high-frequency peak is
suppressed in the $\chi^{''}(Q,\omega)$ spectra, however, it
still is separated from the low-frequency part at the
antiferromagnetic wave vector $Q$. Our results also indicate that
the low-frequency peak of the dynamical susceptibility
$\chi^{''}(Q,\omega)$ is temperature dependent, while the
high-frequency part is almost temperature independent, which are
consistent with the experiments \cite{n5} and numerical simulations
\cite{n9}. The present theoretical results have been used
\cite{n16} to extract the integrated susceptibility and
spin-lattice relaxation time, and the results shown that the
integrated susceptibility exhibits the particularly universal
behavior as
${1/N}\sum_{k}\chi^{''}(k,\omega)\propto {\rm arctan}[a_{1}\omega/T
+a_{3}(\omega/T)^{3}]$ and the spin-lattice relaxation time is
weakly temperature dependent.

In summary, we have studied the spin dynamics of the $t$-$J$
model in the underdoped and optimal doped regimes within the
fermion-spin theory. It is shown that there are two peaks for
the dynamical spin structure factor at the antiferromagnetic wave
vector $Q$, but the high-frequency peak is suppressed in the
dynamical susceptibility spectra $\chi^{''}(Q,\omega)$, which are
qualitative consistent with the experiments and numerical
simulations.

\acknowledgments
This work is supported by the National Science Foundation
Grant No. 19474007, and the Trans-Century Training Programme
Foundation
for the Talents by the State Education Commission of China.

\end{document}